\documentclass[sigconf]{acmart}

\usepackage{tikz}
\usepackage{siunitx}
\definecolor{bluegray}{rgb}{0.4, 0.6, 0.8}
\definecolor{pastelblue}{rgb}{0.68, 0.78, 0.81}
\definecolor{palecornflowerblue}{rgb}{0.67, 0.8, 0.94}
\definecolor{paleaqua}{rgb}{0.74, 0.83, 0.9}
\definecolor{lightcornflowerblue}{rgb}{0.6, 0.81, 0.93}
\definecolor{lightblue}{rgb}{0.68, 0.85, 0.9}
\definecolor{cambridgeblue}{rgb}{0.64, 0.76, 0.68}
\definecolor{ashgrey}{rgb}{0.7, 0.75, 0.71}
\definecolor{beaublue}{rgb}{0.74, 0.83, 0.9}
\definecolor{lightcornflowerblue}{rgb}{0.6, 0.81, 0.93}
\definecolor{moonstoneblue}{rgb}{0.45, 0.66, 0.76}
\definecolor{palecornflowerblue}{rgb}{0.67, 0.8, 0.94}
\definecolor{powderblue}{rgb}{0.69, 0.88, 0.9}

\usepackage{soul, color, colortbl}
\usepackage{amsmath}
\renewcommand\footnotetextcopyrightpermission[1]{}
\usepackage{pifont}
\usepackage{algorithm}
\usepackage{tabularx}
\usepackage{makecell}
\usepackage{multirow}
\usepackage{multicol}
\usepackage{tabularx}
\usepackage{xcolor}
\usepackage{hhline,booktabs,amsmath}
\usepackage[noend]{algpseudocode}
\renewcommand\footnotetextcopyrightpermission[1]{}
\definecolor{mygray}{gray}{0.9}
\definecolor{dkgreen}{rgb}{0,0.6,0}
\definecolor{mauve}{rgb}{0.58,0,0.02}
\definecolor{mygray}{gray}{0.9}
\definecolor{brickred}{rgb}{0.6, 0.0, 0.0}
\newcommand{\sol}{{ThirstyFLOPS}}
\newif\ifdraft
\drafttrue

\ifdraft
  \newcommand{\todo}[1]{{\textcolor{blue}{#1 }}}
  
  \newcommand{\fixme}[1]{{\textcolor{blue}{ Fixme: #1 }}}
\else
  \newcommand{\todo}[1]{}
  \newcommand{\fixme}[1]{}
\fi

\usepackage{xcolor}
\newcounter{takeaway}

\newcounter{TQcounter}

\newcommand{\marconi}{Marconi}
\newcommand{\frontier}{Frontier}
\newcommand{\polaris}{Polaris}
\newcommand{\fugaku}{Fugaku}

\usepackage{xcolor,framed}

\colorlet{TakeawayHeaderBG}{black}
\colorlet{TakeawayHeaderFG}{white}
\colorlet{TakeawayFrame}{black}
\colorlet{TakeawayBodyBG}{gray!10}

\newcommand{\takeaway}[1]{%
\setlength{\fboxsep}{3pt} 
  \par\vspace{6pt}%
  \noindent\colorbox{black}{%
    \makebox[0.955\linewidth][l]{%
      \strut\hspace{10pt}\textcolor{white}{\bfseries Takeaway~\stepcounter{takeaway}\thetakeaway}%
    }%
  }%
  \vspace{-.5mm}
\fcolorbox{black}{lightblue!10}{%
  \parbox{0.95\linewidth}{
    \setlength{\parindent}{0mm}%
    \leftskip=1.2mm
      \rightskip=1.2mm
    \vspace{1.3mm}          
    #1
    \vspace{1.5mm}     
  }%
}
  \par\vspace{3pt}%
}

\usepackage{enumitem}
\setitemize{noitemsep,topsep=0pt,parsep=0pt,partopsep=0pt}
\usepackage[labelfont={sc,bf,it},font+=up,textfont=it,labelsep=space,figurename=Fig.]{caption}

\begin{document}

\title[]{ThirstyFLOPS: Water Footprint Modeling and Analysis\\ Toward Sustainable HPC Systems}

\author{Yankai Jiang}
\affiliation{%
  \institution{Northeastern University}
  \city{Boston}
  \country{USA}}
  \email{jiang.yank@northeastern.edu}

\author{Raghavendra Kanakagiri}
\affiliation{%
  \institution{Indian Institute of Technology Tirupati}
   \city{Tirupati}
  \country{India}}
\email{raghavendra@iittp.ac.in}

\author{Rohan Basu Roy}
\affiliation{%
  \institution{University of Utah}
  \city{Salt Lake City}
  \country{USA}}
\email{rohanbasuroy@sci.utah.edu}

\author{Devesh Tiwari}
\affiliation{%
  \institution{Northeastern University}
    \city{Boston}
  \country{USA}}
  \email{d.tiwari@northeastern.edu}

\begin{abstract}
\textit{High-performance computing (HPC) systems are becoming increasingly water-intensive due to their reliance on water-based cooling and the energy used in power generation. However, the water footprint of HPC remains relatively underexplored—especially in contrast to the growing focus on carbon emissions. In this paper, we present \sol{} - a comprehensive water footprint analysis framework for HPC systems. Our approach incorporates region-specific metrics, including Water Usage Effectiveness, Power Usage Effectiveness, and Energy Water Factor, to quantify water consumption using real-world data. Using four representative HPC systems -- Marconi, Fugaku, Polaris, and Frontier -- as examples, we provide implications for HPC system planning and management. We explore the impact of regional water scarcity and nuclear-based energy strategies on HPC sustainability. Our findings aim to advance the development of water-aware, environmentally responsible computing infrastructures.}
\end{abstract}
\settopmatter{printacmref=false}



\maketitle
\pagestyle{empty}
\renewcommand{\thefootnote}{\fnsymbol{footnote}}
\footnotetext[0]{This paper has been accepted at the 2025 ACM/IEEE International Conference for High Performance
Computing, Networking, Storage, and Analysis (SC '25).}
\renewcommand{\thefootnote}{\arabic{footnote}}

\section{Introduction}
\label{sec:intro}

In recent years, the rapid emergence of energy-intensive workloads, such as generative AI (GenAI)~\cite{cappelloauroragpt}, molecular modeling~\cite{stocks2024breaking}, and climate simulation~\cite{abdulah2024boosting} has driven a dramatic surge in the demand for computational power in HPC systems. Energy consumption has become a crucial challenge for building HPC systems. A recent article pointed out that the HPC systems and datacenters' total electricity consumption could double from 2022 levels to 1,000 terawatt-hours in 2026, approximately the level of electricity demand of Japan~\cite{guardian2025water}.

As the scale and intensity of these workloads continue to grow, so does the energy demand of the underlying HPC infrastructure. This raises urgent questions about the impact of HPC systems on the environment and natural resources. Recent papers have started investigating the environmental impact, in particular the carbon footprint of large-scale systems~\cite{gupta2022act,gupta2021chasing,Acun2023,li2023toward,elgamal2025cordoba,eeckhout2022first, han2025fair,hanafy2024going}, but the water consumption associated with HPC systems remains significantly under-examined compared to carbon metrics.

\vspace{1mm}
\noindent\textbf{Water Footprint of HPC Systems.}
HPC systems consume significant amounts of water. For example, Frontier at Oak Ridge National Laboratory consumes approximately 60 gallons of water per minute~\cite{bloomberg2023frontier,heslin2016ignore}. That is equivalent to 30 million gallons of water per year. To put this in perspective, this is enough water to supply a city of 300 households in the US for an entire year~\cite{daily_water_use}. According to the 2025 sustainability report of Microsoft~\cite{Microsoft2025Sustainability}, the datacenters consumed approximately 1.53 billion gallons of water, equivalent to the annual water usage of more than 10,000 U.S. households~\cite{daily_water_use}. This is more pronounced in water-scarce regions, where the daily household water usage can be very low. For example, rural communities in Ethiopia have an average domestic water consumption of only about 27 gallons per day (an average American family uses more than 300 gallons of water per day at home)~\cite{ahmed2025domestic}. Frontier’s yearly water consumption is much higher by comparison – in fact, one year of Frontier’s water consumption could support 3,000 households in Somalia~\cite{wateractionhub_somalia}.

\vspace{1mm}
Not only do HPC systems consume vast amounts of water, but their impact varies greatly depending on location, creating disproportionate burdens in water-stressed regions where basic needs already compete for limited resources. As noted in WaterWise~\cite{jiang2025waterwise} and other works~\cite{li2023making,wu2025not}, one liter of water in a water-stressed location is more critical than in a water-rich location.

\begin{figure*}[t]
    \centering
\includegraphics[scale=0.48]{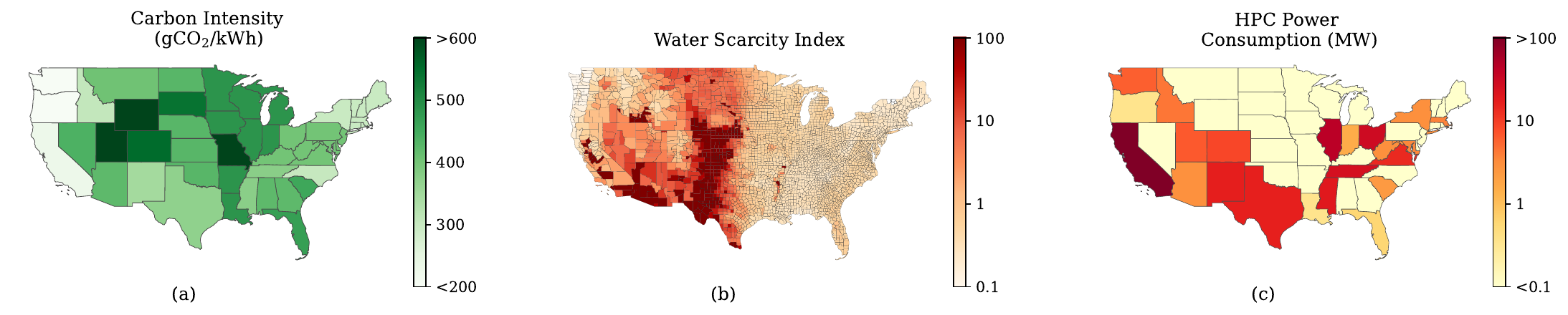}
\vspace{-4mm}
    \caption{Carbon intensity, water scarcity index, and HPC power consumption in the US. 
    }
    \label{fig:heatmap}
    
\end{figure*}

\vspace{2mm}
\textit{To better understand the environmental influence of HPC, we use the US as a case study to provide an overview of different environment-related metrics.} First, we show the carbon intensity in Fig.~\ref{fig:heatmap}(a), the color lightness denotes the magnitude of the carbon intensity, darker green indicates areas with higher carbon intensity (higher carbon intensity is worse for the environment). Note that coastal regions tend to have a slightly lower carbon intensity, and inland regions have a higher carbon intensity. The carbon intensity is collected from Electricity Map~\cite{electricitymap}, since one state may have various power agencies, we only use the carbon intensity number of the major power agencies. 

Next, we show the water scarcity index from AWARE-US~\cite{Lee2018} in Fig.~\ref{fig:heatmap}(b), which reflects the scarcity of the region (refer to Sec.~\ref{sec:background}), darker red areas indicate more severe water scarcity (darker is worse). Finally, we aggregate the power consumption of all US-based supercomputers listed in the HPC TOP500 list~\cite{top500} and visualize the regional HPC power usage in Fig.~\ref{fig:heatmap}(c). 

\vspace{2mm}
Fig.~\ref{fig:heatmap} captures our motivation for this work. \textit{The current HPC centers may not always be located in the most carbon-friendly or water-rich places.} In fact, we observe that some HPC datacenters may be located in relatively water-scarce regions. We recognize that it is not possible or even optimal to place HPC centers in non-water-scarce regions, but currently, we do not systematically consider water consumption, its scarcity, and the carbon footprint of energy sources as main influencing factors when determining where HPC centers should be located and how they should be operated. This is what fundamentally motivates this work. Therefore, we took an operational view of the current HPC practices and asked: \textit{what do we need to do or develop to make our HPC systems more water-efficient?}

As we asked the above question, we recognized that there are a number of gaps -- we simply do not have a systematic method to assess the water footprint of our HPC systems. We acknowledge that the methodology may not be standardized, or the data may not be easily available. But, our HPC community must have the capability to model the water consumption, understand the impact of its different components, their relative importance, and the water footprint's interaction with other sustainability metrics. 

\vspace{2.5mm}
\noindent\textbf{Contribution and Insight Highlights.}

\vspace{2mm}
\noindent\textbf{First, we develop a novel toolset, \sol{}, for estimating the water footprint of HPC systems} -- detailing different components and metrics that should be provided as input to the tool, where to find different data sources to estimate these parameters and their relative importance and contributions (e.g., water consumed during manufacturing stages vs. water consumed during operations). Our \sol{} toolset is open-sourced at~\url{https://doi.org/10.5281/zenodo.15271526} to accelerate the water-aware HPC research.

\vspace{1mm}
\noindent\textbf{Accounting for and comparing the overall water footprint of HPC systems, even during the operational period, is quite involved} -- due to water being consumed indirectly during energy generation and water being consumed directly to cool down the datacenter. \textit{Surprisingly, the indirect water consumption can be comparable to direct water consumption.}

\vspace{1mm}
\noindent\textbf{The water footprint modeling and estimation must account for the relative availability of the water in the region (water scarcity)} -- the utility of the same amount of water volume varies across different geographical locations. Hence, \textit{the water scarcity-unaware site selection of future HPC datacenters can be suboptimal and significantly consequential for future generations (e.g., worsening water shortage in certain locations).}

\vspace{1mm}
\noindent\textbf{This is the first study to reveal that when water is a scarce resource, there are important and difficult decisions to be made} -- \textit{how much of the water should go to cooling the datacenter and how much needs to go toward the generation of electricity that powers the datacenter.} These decisions may need to be coordinated collaboratively between the HPC operators and city power providers. 

\vspace{1mm}
\noindent\textbf{This study demonstrates that different sustainability measures (carbon and water) can be conflicting at times}: Energy sources that are ``greener'' (less carbon footprint) can be, unfortunately, highly water-intensive. Some hardware components that have a lower manufacturing carbon footprint (e.g., hard drives compared to SSDs) can have a high water footprint.

\vspace{1mm}
\noindent\textit{Fortunately, from a programmer's perspective, water-optimized code is essentially equivalent to optimizing for energy optimization.} But, \textbf{the onus of making HPC systems more water-friendly falls on the system software and operations team} -- when to schedule applications for what metric optimization, \textbf{on the facility operators} -- determining what kind of energy to get and from where (significant water scarcity index variation), and \textbf{on the system designers and procurement} -- where to put a new datacenter based on energy and water availability.

\section{Background and Methodology}
\label{sec:background}

\begin{figure*}[t]
    \centering
\includegraphics[scale=0.375]{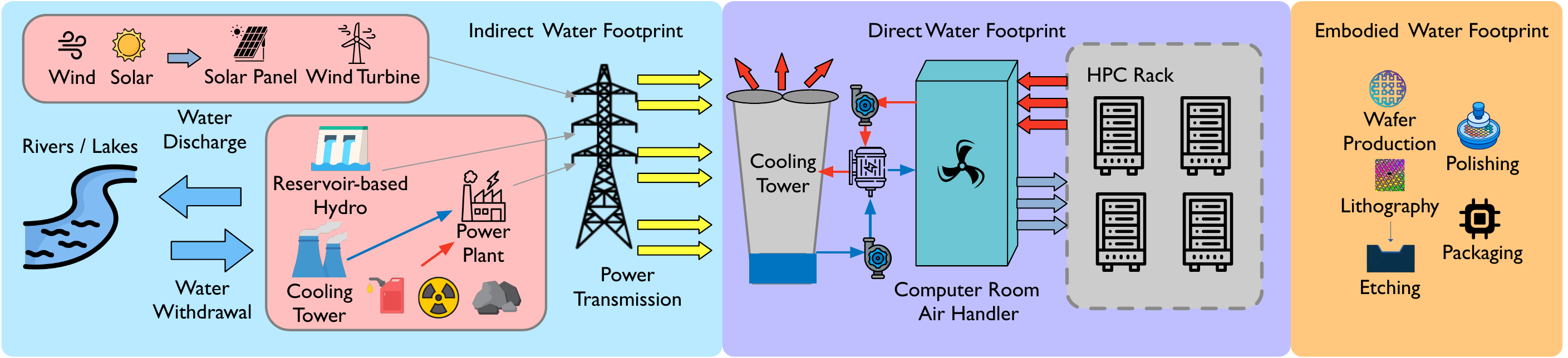}
    \caption{Water footprint components in the HPC systems, including operational water footprint and embodied water footprint.
    }
    \label{fig:water_hpc}
\end{figure*}

In this section, we discuss the different phases of water consumption and their corresponding purposes (e.g., manufacturing and operating an HPC system). As we define and describe these various components of an HPC system's water footprint, we also provide formulations and relationships among them. Then, we discuss the methodology to estimate these water footprint components. 

First, we note the distinction between water withdrawal and consumption: withdrawal refers to the total amount of water removed from a source (like a river or lake)~\cite{reig2013s,cohen2014water}, while consumption is defined as ``water withdrawal minus water discharge'', which represents the portion of that water which is evaporated or otherwise permanently removed from the immediate water environment~\cite{siddik2021environmental}. In this work, water footprint refers to the water consumption during manufacturing and operating the HPC systems~\cite{li2023making}. 

In Fig.~\ref{fig:water_hpc}, we visualize different water footprint components in the HPC system. The HPC system has two major water footprint components: \emph{embodied water footprint} ($W_{\text{embodied}}$) and \emph{operational water footprint} ($W_{\text{operational}}$). The embodied water footprint is a one-time consumption, and it accounts for the water used in hardware manufacturing, transportation, and disposal (the rightmost component in Fig.~\ref{fig:water_hpc}). We note that the majority of the embodied water footprint comes from the manufacturing site and not necessarily where the HPC datacenter is located. 

The operational water footprint is the water consumed during the operations of HPC systems (i.e., during application execution on the HPC system to power and cool the HPC system). It has two components \emph{direct water footprint} ($W_{\text{direct}}$) and \emph{indirect water footprint} ($W_{\text{indirect}}$)~\cite{li2023making,jin2019water}. The direct water footprint refers to the water consumption inside the datacenter to cool the datacenter (the middle component in Fig.~\ref{fig:water_hpc}). The indirect water footprint refers to the water consumption outside the datacenter during the energy generation process -- energy that is used to power the datacenter (the leftmost component in Fig.~\ref{fig:water_hpc}). The total water footprint ($W$) can be formulated as Eq.~\ref{eq:overall_model}. In the next section, we model each water footprint component.

\vspace{-1em}
{\small
\begin{align}
\begin{split}
\label{eq:overall_model}
W=W_{\text{embodied}}+W_{\text{operational}}=W_{\text{embodied}}+W_{\text{direct}}+W_{\text{indirect}}
\end{split}
\end{align}}
\vspace{-5mm}

\subsection{Embodied Water Footprint Modeling}

\label{sec:emb_model}
\noindent\textbf{Embodied Water Footprint.} 
The embodied water footprint is the water footprint generated during hardware manufacturing -- namely, during packaging and manufacturing. We leverage similar modeling approaches from carbon modeling works~\cite{gupta2022act,ji2024scarif} to divide the embodied water footprint ($W_{\text{embodied}}$) into the packaging water footprint ($W_{\text{pkg}}$) and the manufacturing water footprint ($W_{\text{mfg}}$). 

\vspace{-1em}
{\small
\begin{align}
\begin{split}
\label{eq:embodied_model}
W_{\text{embodied}}=W_{\text{pkg}}+W_{\text{mfg}}
\end{split}
\end{align}}
\vspace{-1em}

\noindent The packaging water footprint refers to the total water consumption of packaging the integrated circuits of different hardware (processors, memory, and storage devices). The packaging water footprint of one hardware component is the product of the water footprint overhead per integrated circuit ($W_{\text{IC}}$) and the number of integrated circuits ($N_{\text{IC}}$). So the $W_{\text{pkg}}$ is expressed in Eq.~\ref{eq:pacakge_model}.

\vspace{-1em}
{\small
\begin{align}
\begin{split}
\label{eq:pacakge_model}
W_{\text{pkg}}=\sum_{\text{IC}}^{\text{devices}}W_{\text{IC}}\times N_{\text{IC}}
\end{split}
\end{align}}

The manufacturing water footprint ($W_{\text{mfg}}$) is modeled in two steps - first for the processors, and then, for the memory and storage. The manufacturing water footprint for processors (e.g., CPU and GPU) can be modeled via the processes from the architecture technology perspective. During the manufacturing process of processors, large amounts of ultrapure water (UPW) are used in silicon wafer production, lithography, and etching, to clean and remove impurities or contaminants. Then, the following step is chemical mechanical polishing, which requires process cooling water (PCW) to cool and rinse (one of the most water-intensive processes in chip manufacturing). Additionally, the manufacturing process requires energy to power the whole process, which requires water during the energy generation process. This water usage is quantified as the water required to power each unit of die area (WPA). We formulate the manufacturing water footprint of processors in Eq.~\ref{eq:man_processor}.

\vspace{-1em}
{\small
\begin{align}
\begin{split}
\label{eq:man_processor}
W_{\text{mfg}}^{\text{CPU, GPU}}=\frac{1}{\text{Yield}}\cdot A_{\text{die}}(\text{UPW}+\text{PCW}+\text{WPA})
\end{split}
\end{align}}

\noindent Here, $A_{\text{die}}$ is the die area of the chips. Yield is the fab yield rate. 

Finally, for the memory and storage devices, the manufacturing water footprint can be modeled as the product of the water footprint per capacity (WPC) and the capacity of the memory and storage devices (Eq.~\ref{eq:man_mem}).

\vspace{-1em}
{\small
\begin{align}
\begin{split}
\label{eq:man_mem}
W_{\text{mfg}}^{\text{DRAM, SSD, HDD}}=\text{WPC}\cdot\text{Capacity}
\end{split}
\end{align}}

We note that accurate modeling of the embodied water footprint relies on multiple different parameters described above, including water footprint overhead per integrated circuit ($W_{\text{IC}}$), the number of integrated circuits ($N_{\text{IC}}$), die area of the chips ($A_{\text{die}}$), Yield, UPW, PCW, WPA, WPC, and capacity of memory/storage devices. Fortunately, these parameter values are provided in vendor sheets, and can be estimated via manufacturing sites and processor technology as described in Sec.~\ref{sec:method}.

\subsection{Operational Water Footprint Modeling}
\label{sec:op_model}

The operational water footprint is categorized into two components: direct water footprint and indirect water footprint. Recall that the operational water footprint is water consumption during the operations of the HPC center -- and some parts occur on the HPC facility site (to cool the datacenter) referred to as direct water footprint, and some parts occur away from the HPC facility site, but nearby where the energy is being generated to power the HPC center (referred to as indirect water footprint). 

\vspace{1mm}
\noindent\textbf{Direct Water Footprint.} \textit{The direct water footprint is consumed during the cooling process in the HPC systems.} The computer room air handler uses chilled water from the chiller to cool the air in the HPC rack rooms and uses fans to maintain a set temperature and humidity. The chilled water generated from the cooling tower is evaporated or lost via blowdown. This amount of water is the direct water footprint. The direct water footprint is computed as the product of the energy consumed by running the HPC application ($E$) and the water usage effectiveness (WUE) during the use phase as Eq.~\ref{eq:direct}. \textit{WUE (measured in L/kWh) quantifies the amount of water needed to cool one unit of energy (lower is better), and it is determined by the locational outside wet bulb temperature~\cite{li2023making,islam2015water}, which is a function based on air temperature and humidity~\cite{stull2011wet}.}

\vspace{-1em}
{\small
\begin{align}
\begin{split}
\label{eq:direct}
W_{\text{direct}} &= E\cdot\text{WUE}\\
\text{WUE} &=\frac{\text{cooling water}}{\text{power consumption} }= f(\text{Air}_{\text{temperature}}, \text{humidity})
\end{split}
\end{align}}

Intuitively, if an HPC system executes applications that are very power-intensive (compute-bound applications and accelerator-heavy applications) and the HPC system utilization is high, then, the direct water footprint will be higher because of high energy demand. The direct water footprint also depends on the water usage effectiveness of the HPC facility -- which in turn, depends on the location of the datacenter. If the HPC facility is located in a favorable geographical location or time of the year, the outside air can be used for cooling the datacenter, and hence, the water footprint is expected to be lower. \textit{That is, a favorable outside climate leads to a lower water footprint and lower WUE. As expected, WUE can vary across geographical locations and even temporally for a single geographical location -- even when the HPC system, workloads, and facility capacity may be the same.}

\vspace{1mm}
\noindent\textbf{Indirect Water Footprint.} HPC systems rely on electric power, which is generated using different energy sources -- often, using a mix of energy sources (e.g., fossil fuel and renewable energy). \textit{The indirect water footprint refers to the amount of water consumed during energy generation. The energy water factor (EWF) is the metric that determines the indirect water footprint. Different energy sources have different EWFs, a small EWF indicates a small amount of water consumed when transforming energy sources into electricity.} Note that some renewable energy sources, such as geothermal, nuclear, and hydro power, exhibit higher EWF compared with fossil energy, such as coal, gas, and oil~\cite{macknick2011review,reig2020guidance}. The EWF of a specific region is calculated by the weighted summing of the EWFs of the energy mix used in the region.

The indirect water footprint is calculated by multiplying the effective energy usage of the HPC systems by the regional EWF. Specifically, the energy usage of the HPC systems is determined by the power usage effectiveness (PUE) and the energy consumption ($E$). PUE quantifies the energy efficiency of the HPC system and datacenters by comparing the total facility energy to the energy used directly by IT equipment. A lower PUE indicates higher efficiency, with a value of 1 representing an ideal case where all consumed energy powers the IT equipment only. In practice, modern supercomputers can achieve relatively low PUE values ~\cite{shehabi2016united,shin2021revealing}. Combining this efficiency metric, Eq.~\ref{eq:indirect} shows the formula for indirect water footprint calculation.

\vspace{-1em}
{\small
\begin{align}
\begin{split}
\label{eq:indirect}
W_{\text{indirect}} &= E\cdot\text{PUE}\cdot \text{EWF}\\
\text{EWF}&=\frac{\text{energy generation water}}{\text{power generated}}=f(\text{mix}\%, \text{EWF}_{\text{energy}})
\end{split}
\end{align}}

\noindent\textbf{Water Intensity (WI).} The energy consumption ($E$) can be extracted from the total operational water footprint ($W_{\text{direct}}+W_{\text{indirect}}$), and the remaining part can be defined as water intensity (WI). \textit{Water intensity can simplify the calculation of water consumed during HPC operations and can act as a proxy for water footprint for the operational water footprint - similar to carbon intensity for carbon footprint.} WI considers both direct and indirect parts, and water intensity is defined as the following equation.

\vspace{-1em}
{\small
\begin{align}
\begin{split}
\label{eq:wi}
W_{\text{operational}} &= W_{\text{direct}}+W_{\text{indirect}}\\&=E\cdot\text{WUE}+E\cdot\text{PUE}\cdot\text{EWF}\\ &=  E\cdot(\text{WUE}+\text{PUE}\cdot\text{EWF})=E\cdot\text{WI}\\
\text{Water Intensity (WI)} &= \text{WUE}+\text{PUE}\cdot\text{EWF}
\end{split}
\end{align}}

Here, the WUE and $\text{PUE}\cdot\text{EWF}$ refer to the direct and indirect water intensity ($\text{WI}^{\text{direct}}$ and $\text{WI}^{\text{indirect}}$).

\vspace{1mm}
\noindent\textbf{Regional Water Scarcity (WSI).} Finally, we note that when calculating operational water footprints, it is critical to consider local water stress levels, as the impacts of water consumption vary based on regional scarcity. \textit{The water scarcity index (WSI) accounts for geographic variation by applying weighting factors to volumetric water use~\cite{Lee2018,chen2023sustainability}. A higher WSI indicates that the region is more water-stressed.} By scaling the operational water footprint with the WSI, a ``water scarcity-aware water footprint'' is produced, clearly illustrating how water consumption contributes to resource depletion in specific areas. Eq.~\ref{eq:wsf} shows the new water intensity after WSI adjustment. Several methods~\cite{Falkenmark1989,Raskin1997,Stefano2020,Pfister2009,wu2025not} have been developed to measure regional water stress. Early approaches provided broad estimates suitable for large areas.  
{\small
\begin{align}
\vspace{-3mm}
\begin{split}
\label{eq:wsf}
\text{WI}^{\text{WSI}} &= \text{WI}\cdot\text{WSI}
\end{split}
\end{align}}

\begin{table}
\centering
\caption{Supercomputers used in water footprint analysis. }
\vspace{-2mm}
\scalebox{0.62}{
\begin{tabular}{cccc}
\toprule
\textbf{Name} & \textbf{Location}& \textbf{Processor (CPU/GPU)} &\textbf{Start Year}\\ 
\midrule
\midrule
\textbf{\marconi{}~\cite{marconi100}} &  \makecell[c]{Bologna, Italy\\CINECA}
&  \makecell[c]{IBM Power9 AC922\\NVIDIA V100 SXM2}
&\makecell[c]{2019}\\
\midrule
\textbf{\fugaku{}~\cite{fugaku}} &  \makecell[c]{Kobe, Japan\\Riken CCS}
&  \makecell[c]{Fujitsu A64FX 48C\\ No GPU} 
&\makecell[c]{2020}\\
\midrule
\textbf{\polaris{}~\cite{polaris}} &\makecell[c]{Lemont, IL, US\\Argonne National Lab} &\makecell[c]{AMD EPYC 7532\\NVIDIA A100 PCIe} 
&\makecell[c]{2021}\\
\midrule

\textbf{\frontier{}~\cite{frontier}} &  \makecell[c]{Oak Ridge, TN, US \\Oak Ridge National Laboratory} 
&  \makecell[c]{AMD EPYC 7A53 \\AMD Instinct MI250X} &\makecell[c]{2021}\\
\bottomrule
\end{tabular}}
\label{table:info}
\vspace{-4mm}
\end{table}

\begin{table*}[t]
\centering
\caption{Parameters for estimating the operational and embodied water footprint for the HPC system.}
\vspace{-3mm}
\scalebox{0.62}{
\begin{tabular}{p{0.3cm}ccccccc}
\toprule
\textbf{} & \textbf{Parameter}&\textbf{Parameter Description} &\textbf{Input \ding{109}/Derive \ding{115}}&\textbf{Data Range}&\textbf{Data Source} &\textbf{Unit} &\textbf{Reference} \\  
\midrule
\bottomrule

\multirow{11}{*}{\rotatebox{90}{$W_{\text{embodied}}$}} & \makecell[c]{$N_{\text{IC}}$} &\makecell[c]{Number of ICs (CPU/GPU/memory/storage)} &  \makecell[c]{\ding{109}}&\makecell[c]{9-26 (Vary across hardware)}&\makecell[c]{From hardware design}&\makecell[c]{None}&\makecell[c]{\cite{gupta2022act,li2023toward}}\\

\cline{2-8} & \makecell[c]{$W_{\text{IC}}$}& \makecell[c]{Packaging water overhead}& \makecell[c]{\ding{115}}&\makecell[c]{0.6}&\makecell[c]{From manufacturer}&\makecell[c]{$L$}&\makecell[c]{\cite{spil2019sustainability,Thum2022}}\\

\cline{2-8} & \makecell[c]{$A_{\text{die}}$}& \makecell[c]{Die size of processors (CPU/GPU)}& \makecell[c]{\ding{109}}&\makecell[c]{Vary across hardware}&\makecell[c]{From CPU/GPU design}&\makecell[c]{$mm^2$}&\makecell[c]{\cite{nohria2018ibm,techpowerup,matsuoka2021fugaku}}\\ 

\cline{2-8} & \makecell[c]{Yield}& \makecell[c]{Fab yield rate of hardware manufacturing}&\makecell[c]{\ding{109}}&\makecell[c]{0-1 (0.875 as default)}&\makecell[c]{From manufacturer}&\makecell[c]{None}&\makecell[c]{\cite{gupta2022act}}\\

\cline{2-8} & \makecell[c]{Location}& \makecell[c]{Manufacturing location of hardware}& \makecell[c]{\ding{109}}& \makecell[c]{TSMC or GlobalFoundries}& \makecell[c]{From manufacturer}&\makecell[c]{None}&\makecell[c]{\cite{wikichip}}\\ 

\cline{2-8} & \makecell[c]{Process Node}& \makecell[c]{Semiconductor manufacturing process of CPU/GPU}& \makecell[c]{\ding{109}}& \makecell[c]{3-28 (Vary across hardware)}& \makecell[c]{From CPU/GPU design}& \makecell[c]{$nm$}& \makecell[c]{\cite{bardon2020dtco}}\\

\cline{2-8} & \makecell[c]{UPW}& \makecell[c]{Ultrapure water usage during manufacturing}& \makecell[c]{\ding{115}}& \makecell[c]{5.9-14.2 (Vary across process node)}& \makecell[c]{From manufacturer}& \makecell[c]{$L$}& \makecell[c]{~\cite{bardon2020dtco}}\\

\cline{2-8} & \makecell[c]{PCW}& \makecell[c]{Process cooling water during manufacturing}& \makecell[c]{\ding{115}}& \makecell[c]{Vary across locations and process node}& \makecell[c]{From manufacturer}& \makecell[c]{$L$}& \makecell[c]{~\cite{bardon2020dtco}}\\

\cline{2-8} & \makecell[c]{WPA}& \makecell[c]{Water for power generation during manufacturing}& \makecell[c]{\ding{115}}& \makecell[c]{Vary across locations and process node}& \makecell[c]{From manufacturer}& \makecell[c]{$L$}& \makecell[c]{~\cite{bardon2020dtco}}\\

\cline{2-8} & \makecell[c]{WPC}& \makecell[c]{Water footprint per capacity of DRAM, HDD, SSD}& \makecell[c]{\ding{115}}& \makecell[c]{0.8 (DRAM), 0.033 (HDD), 0.022(SSD)}& \makecell[c]{From manufacturer}& \makecell[c]{$L/\text{GB}$}& \makecell[c]{\cite{seagate_hdd,seagate_ssd,SKhynix2021}}\\

\cline{2-8} & \makecell[c]{Capacity}& \makecell[c]{Capacity of DRAM, HDD, SSD}& \makecell[c]{\ding{109}}& \makecell[c]{Vary across hardware}& \makecell[c]{From manufacturer}& \makecell[c]{GB}& \makecell[c]{\cite{frontier,marconi100,fugaku,polaris}}\\

\midrule
\midrule

\multirow{9}{*}{\rotatebox{90}{$W_{\text{operational}}$}} & \makecell[c]{$E$} & \makecell[c]{Energy consumption} & \makecell[c]{\ding{109}}& \makecell[c]{Vary across applications/hardware}& \makecell[c]{From hardware profiling}& \makecell[c]{$kWh$}& \makecell[c]{\cite{treibig2010likwid,solorzano2024toward,sun2024energy}}\\

\cline{2-8} & \makecell[c]{Wet bulb temperature}& \makecell[c]{Site-related wet bulb temperature}&\makecell[c]{\ding{109}}&\makecell[c]{Vary across HPC locations}&\makecell[c]{From weather report}&\makecell[c]{${^{\circ}C}$}&\makecell[c]{\cite{meteologix}}\\

\cline{2-8} & \makecell[c]{WUE}& \makecell[c]{Water usage effectiveness}&\makecell[c]{\ding{115}}&\makecell[c]{>0.05}&\makecell[c]{From wet bulb temperature}&\makecell[c]{$L/kWh$}&\makecell[c]{\cite{gupta2024dataset}}\\

\cline{2-8} & \makecell[c]{PUE}& \makecell[c]{Power Usage Effectiveness}&\makecell[c]{\ding{109}}&\makecell[c]{$\geq$1 (Marconi: 1.25, Fugaku:1.4, \\Polaris:1.65, Frontier: 1.05 )}&\makecell[c]{From HPC report}&\makecell[c]{None}&\makecell[c]{\cite{cineca:ppt,terai2020study,2020argonne,sun2024energy}}\\

\cline{2-8} & \makecell[c]{mix\%}& \makecell[c]{Percentage energy mix usage}&\makecell[c]{\ding{109}}&\makecell[c]{0-100}&\makecell[c]{From power grid}& \makecell[c]{\%}& \makecell[c]{\cite{electricitymap}}\\ 

\cline{2-8} & \makecell[c]{EWF$_{\text{energy}}$}& \makecell[c]{energy water factor of energy sources}&\makecell[c]{\ding{115}}&\makecell[c]{1-17}&\makecell[c]{From environment report}& \makecell[c]{$L/kWh$}& \makecell[c]{\cite{reig2020guidance,macknick2011review}}\\

\cline{2-8} & \makecell[c]{EWF}& \makecell[c]{energy water factor of HPC system}&\makecell[c]{\ding{115}}&\makecell[c]{Vary across locations}&\makecell[c]{From mix\% and EWF$_{\text{energy}}$}& \makecell[c]{$L/kWh$}& \makecell[c]{\cite{electricitymap}}\\

\cline{2-8} & \makecell[c]{WSI$^{\text{direct}}$}& \makecell[c]{Direct water scarcity index}& \makecell[c]{\ding{109}}& \makecell[c]{0.1-100}& \makecell[c]{From WSI report}&\makecell[c]{None}& \makecell[c]{\cite{Lee2018,Boulay2018,aqueduct}}\\ 

\cline{2-8} & \makecell[c]{WSI$^{\text{indirect}}$}& \makecell[c]{Indirect water scarcity index}& \makecell[c]{\ding{109}}& \makecell[c]{0.1-100}& \makecell[c]{From WSI report and power plant locations}& \makecell[c]{None}& \makecell[c]{\cite{Lee2018,Boulay2018,aqueduct}}\\

\bottomrule
\end{tabular}}

\label{table:tool}
\end{table*}

In this work, by default, we use the water footprint without explicitly incorporating the water scarcity index to decouple the effects of water and its regional scarcity. But, we also separately evaluate the impact of WSI on the overall water footprint.

\subsection{Methodology}
\label{sec:method}

\vspace{1mm}
\noindent\textbf{Systems.} We use four supercomputers as a case study for this work: \marconi{}100~(Italy), \fugaku{}~(Japan), \polaris{}~(US), and \frontier{}~(US) across different countries and computational power (listed in HPC TOP500 list~\cite{top500} at various points).
We note that we have selected them to represent a diverse set of choices, but we do not claim that they cover the full spectrum. Our intent is to show that, by using our tool, facility designers, operators, and practitioners can learn various trade-offs and draw useful comparisons. 

\begin{figure}[t]
    \centering
\includegraphics[scale=0.40]{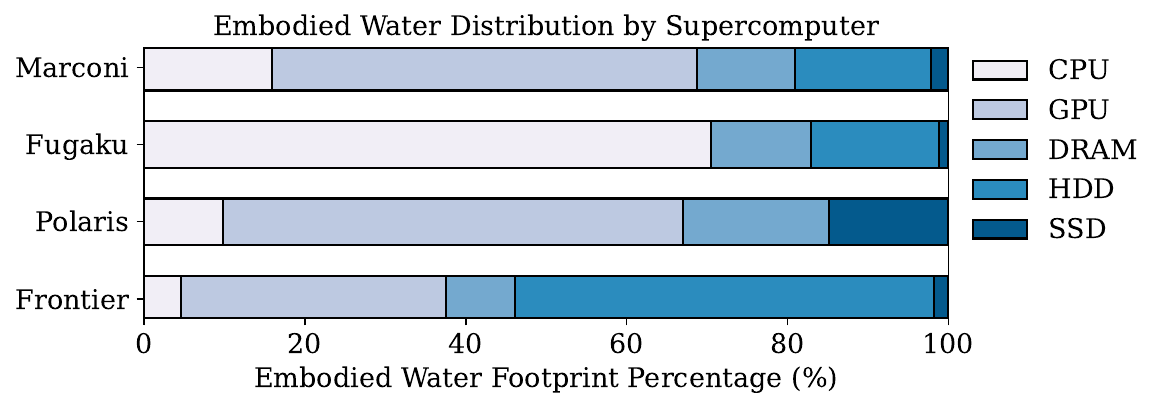}
\vspace{-3mm}
    \caption{Embodied water footprint contribution of different hardware components, including CPU, GPU, DRAM, HDD, and SSD.}
    \label{fig:emb_water}
\vspace{-4mm}
\end{figure}

The detailed hardware configurations for these four supercomputers are summarized in Table~\ref{table:info}. Our analysis includes both the embodied water footprint of the hardware and the regional water resource characteristics at their respective locations — Bologna (Italy), Kobe (Japan), Lemont (US), and Oak Ridge (US). We also examine machine-level job utilization and power usage based on available system logs. Specifically, we use 2021–2022 log data of \marconi{}~\cite{borghesi2023m100}, 2023 log data of \polaris{}~\cite{ALCFPublicData}, 2023-2024 log data of \fugaku{}~\cite{solorzano2024toward}, and 2023 power log data of \frontier{}~\cite{sun2024energy}. These monitors record the jobs running on the HPC system. If power consumption data is available, we use it directly; otherwise, we calculate the machine utilization from job logs and estimate the energy consumption of the supercomputer using the hardware's thermal design power.

\vspace{2mm}
\noindent\textbf{Water Footprint Estimation Tool.} Table~\ref{table:tool} summarizes the input parameters and corresponding data references used in our modeling tool -- \sol{}, to estimate both the embodied and operational water footprint of HPC systems -- \textit{\sol{} is the first-of-its-kind framework and is open-sourced to the community}. In particular, we describe all the parameters that are required for different calculations, their corresponding data sources, and expected data ranges. Before using our tool, the HPC datacenter researchers can use this table as a checklist to know what information is required and pointers on where to get it, and which part of the estimations they contribute toward. Specifically, the embodied water footprint ($W_\text{embodied}$) of various hardware components is estimated following the methodology described in Sec.\ref{sec:emb_model}, while the operational water footprint ($W_\text{operational}$) is computed using water intensity values and HPC power consumption as outlined in Sec.\ref{sec:op_model}. 

We emphatically acknowledge that due to the infancy stage of water footprint modeling and lack of standardization, the modeling tool will continue to be enhanced with community input and vendor information. To account for that, in our evaluation, we focus on comparative trade-offs, and trends instead of claiming typical \%-based improvement by deploying a particular strategy. We also explicitly acknowledge when a particular result may be susceptible to such unavoidable estimation differences. 

\section{Water Footprint Analysis}

In this section, we first show a detailed breakdown of the embodied water footprint for each hardware component in HPC systems. Next, we compare the direct and indirect water footprints.

\vspace{2mm}
\noindent\textbf{\large{Embodied Water Footprint.}} In Fig.~\ref{fig:emb_water}, we show a comparative analysis of the embodied water footprint contributions from key hardware components, processors (CPU and GPU), memory (DRAM), and storage (SSD and HDD). From the figure, we make two observations. First, as expected, GPUs contribute significantly to the overall embodied water footprint for GPU-rich systems (e.g., \frontier{}, \marconi{}, and \polaris{}). For example, in \polaris{} - using the A100 PCIe 40GB GPU, GPUs account for 67\% of its total embodied water footprint. We do not purposely compare the absolute water footprint of GPU vs. CPU because it is dependent on multiple factors, including technology generation, performance/water ratio, etc. Nevertheless, as expected, computing components can be responsible for a relatively significant fraction of the embodied water footprint compared to other components.

\vspace{1mm}
Second, interestingly, although the embodied water footprint of memory and storage components is generally lower than that of processors, their relative contribution can become more significant in certain systems. For example, in \frontier{}, which features a 679~PB HDD-based file system, its overall storage and memory embodied water footprint is 24.8\% point higher than that of its processors (CPUs and GPUs). In \marconi{}, \fugaku{} and \polaris{}, memory and storage components account for 27\% of the total embodied water footprint for all three systems. Notably, \polaris{} employs an all-flash storage, which substantially reduces the water footprint associated with storage compared to HDD-based systems. 

The primary reason for the above finding is that HDDs include materials like lubricants, adhesives, rare-earth magnets, and precious metals in their PCB—many of which require significant water use during extraction, cooling, or chemical processing~\cite{epa2025standards}. In contrast, an SSD has a simpler bill of materials, primarily consisting of silicon, plastics, and a small amount of metal on the PCB. SSDs contain significantly lower quantities of certain metals compared to HDDs~\cite{kim2019environmental}. These factors lead to lower water footprints in SSDs. Next, we naturally investigate the relative contributions of embodied and operational components toward the overall water footprint.

\takeaway{
The HPC systems which have a large storage capacity backed by traditional hard disk drives, have a significant embodied water footprint coming from hard drives. SSDs, while more expensive, are favorable in terms of embodied water footprint compared to hard drives. This trade-off is the exact opposite when we consider only the embodied carbon footprint~\cite{li2023toward,tannu2023dirty}, where SSDs are reported to have a higher embodied carbon footprint than hard drives. \textit{Achieving practical environmental sustainability of an HPC system is challenging for facility designers -- different HPC components rank differently on different sustainability metrics (carbon vs. water).}  
}

\begin{figure}[t]
    \centering
\includegraphics[scale=0.41]{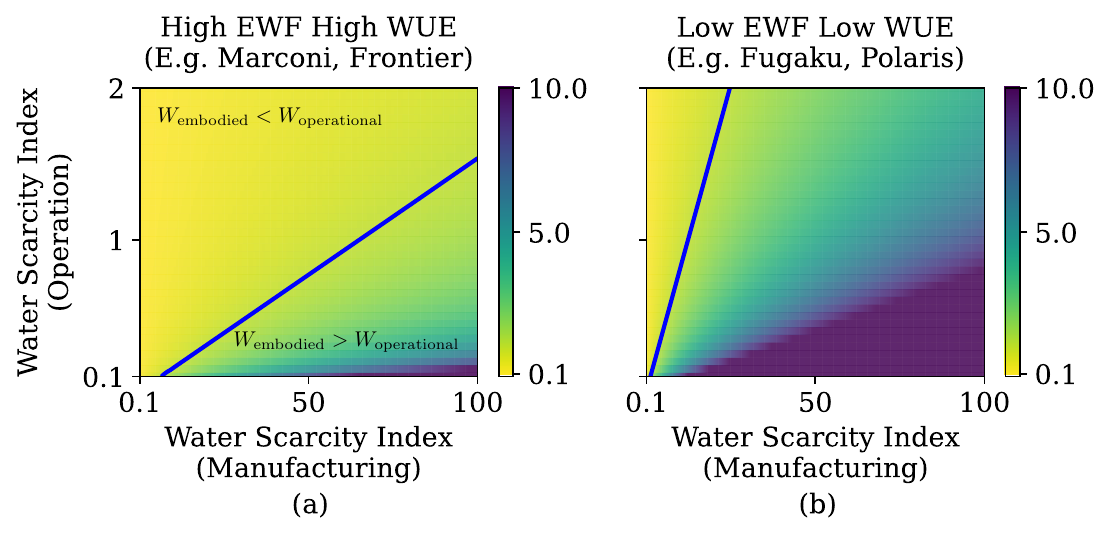}
\vspace{-4mm}
    \caption{Comparison of embodied and operational water footprint under different EWF, WUE, and WSI scenarios. 
    }
    \vspace{-4mm}
    \label{fig:heat}
\end{figure}

\vspace{2mm}
\noindent\textbf{\large{Embodied vs Operational Water Footprint.}} Recall that operational water footprint depends on two key factors: EWF metric that captures the water consumption during energy generation (which depends on the energy mix), and WUE metric that captures the weather conditions in the datacenter location (affects the water needed to cool down the system). Furthermore, the water footprint is influenced by regional water scarcity indexes (WSIs) -- how scarce the water is, as a resource. 

Fig.~\ref{fig:heat} visualizes the comparisons between embodied and operational water footprints under different scenarios. Broadly, we attempt to understand what happens when the water scarcity index changes, and when the EWF and WUE change. Fig.~\ref{fig:heat} shows two representative scenarios: case (a) with high EWF and high WUE, and case (b) with low EWF and low WUE. 

We calculate the ratio of embodied to operational water footprint ($\frac{W_\text{embodied}}{W_\text{operational}}$) to determine which component dominates under different scenarios. A blue line is used to indicate the boundary where the ratio equals 1, which means $W_\text{embodied}={W_\text{operational}}$. The region below the blue line represents cases where the embodied water footprint exceeds the operational footprint in HPC systems. The color shading in the heatmap reflects the magnitude of the ratios, darker colors indicate higher ratios, meaning a more dominant embodied water footprint. Below, we summarize the findings.

If the HPC systems are built in a water-secure region (low operational WSI), while the hardware is manufactured in a water-scarce region (high manufacturing WSI), the embodied water footprint can exceed the operational water footprint. 

If the EWF and WUE of the HPC system are high (case (a); that is water-intensive energy generation and significant water needed to cool the datacenter due to unfavorable weather conditions), the area below the blue line becomes smaller, indicating that the embodied water footprint is less likely to reach or exceed the operational water footprint. In contrast, when the EWF and WUE are low (case (b); that is less water-intensive energy generation and favorable weather conditions for cooling the datacenter), the area below the blue line expands, suggesting that the embodied water footprint more easily surpasses the operational component.

\begin{figure}[t]
    \centering
\includegraphics[scale=0.4]{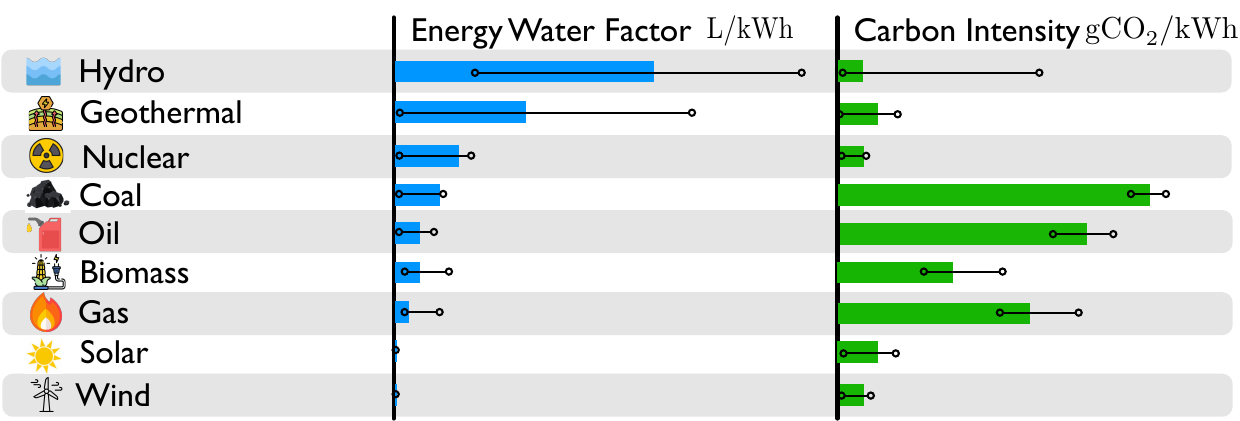}
\vspace{-3mm}
    \caption{Different energy sources have different energy water factors (EWFs) and carbon intensities.
    }
    \label{fig:diff_energy}
\end{figure}

\takeaway{
Both the geographic location of the hardware manufacturer and the geographic location of the HPC center play a critical role in the system's overall water footprint. Building fabrication facilities in water-scarce regions can lead to disproportionately high embodied water footprints, even if operational water use remains low. \textit{Therefore, careful consideration of manufacturing sites is critical for HPC systems, in addition to operational site selection for HPC systems~\cite{hpcwire2025doe}. Even for operational site selection, we highlight the need for modeling and accounting for the water intensity of energy generation, year-round weather conditions, and water scarcity of the local region.}
}

\vspace{2mm}
\noindent\textbf{\large{Direct and Indirect Operational Water Footprint.}} Next, we investigate the operational water footprint in greater depth. Recall from Sec.~\ref{sec:background} that the operational water footprint comprises direct and indirect water footprints. As discussed earlier, two factors (EWF and WUE) affect these components. EWF (water intensity of the energy generation process) influences the indirect water footprint, and WUE (climate conditions in the datacenter region) influences the direct water footprint. Therefore, first, we analyze these two factors.

\begin{figure}[t]
    \centering
\includegraphics[scale=0.40]{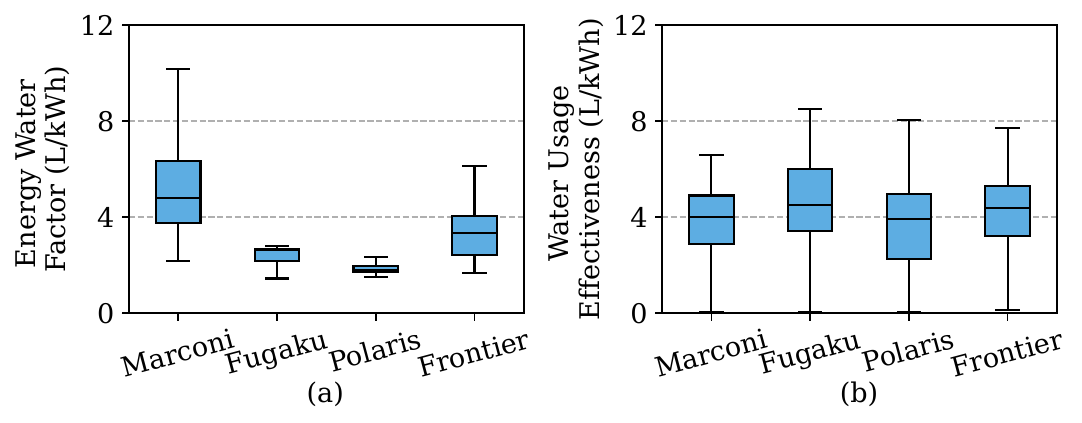}
\vspace{-4mm}
    \caption{Water consumption during energy generation (EWF) and cooling (WUE) has significant temporal and spatial variation.}
    \label{fig:ewf_wue_variation}
    \vspace{-5mm}
\end{figure}

\begin{figure}[t]
    \centering
\includegraphics[scale=0.44]{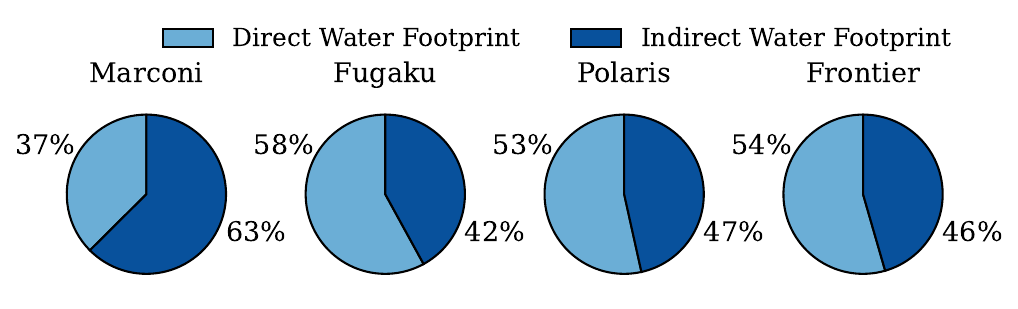}
\vspace{-3.5mm}
    \caption{Relative importance of direct and indirect water footprint.}
    \vspace{-4mm}
    \label{fig:indirect_vs_direct}
\end{figure}

First, Fig.~\ref{fig:diff_energy} shows the EWF (a measure of water consumption during energy generation) for different energy sources (e.g., hydro, biomass, solar, wind, etc.) and carbon intensity (a measure of $CO_2-eq$ emission) of these energy sources. The bar represents the median value, while the black line indicates the range between the minimum and maximum values. Interestingly, we note that energy sources that are ``greener'' (lower carbon intensity) may require significant water during energy generation (e.g., hydro and geothermal). Note that, our EWF values for hydroelectricity reflect aggregated in-stream and reservoir (primarily dominated) data, implicitly including evaporation losses. The EWF of hydroelectric power varies depending on the mix of in-stream and reservoir shape (depth and width)~\cite{scherer2015water}. For example, a wide but shallow reservoir may lead to high evaporation (relatively high EWF, compared to less wide and deeper reservoirs), leading to observed variation.

Each HPC facility may use a mix of energy sources and the availability of energy sources may change or have significant seasonal variation (e.g., 30\% solar and 70\% coal to 50\% biomass and 50\% coal) -- this is reflected in Fig.~\ref{fig:ewf_wue_variation} (a), where each facility observes different EWF and significant variation in EWF over the year. For example, \marconi{} has the widest variation range. The wide range is primarily due to its reliance on hydro power. In particular, the availability of hydro power fluctuates over time, which has a pronounced impact on the EWF, given the high EWF in hydro. The EWF in \marconi{} can become 10.59~L/kWh, the highest among all evaluated regions. In contrast, \polaris{} has the lowest EWF, which can reach 1.52~L/kWh, 85\% lower than \marconi{}'s.

\takeaway{Water consumption during electricity generation can be significant, and energy sources that are typically environment-friendly (in terms of $CO_2$-$\text{eq}$ emissions) are not necessarily water-friendly and their water consumption can have more than 50\% variation temporally. 
}

\vspace{1mm}
Second, Fig.~\ref{fig:ewf_wue_variation} (b) shows that even the water consumption on the datacenter facility to cool down the hardware can have even wider temporal variation (reflected in the WUE factor) -- due to changes in the outside humidity and wet-bulb temperature. While this variation is expected, we note that the scale of the change in WUE and EWF is in a similar range. This indicates that components are important to the overall operational water footprint (the two additive factors in Eq.~\ref{eq:wi}). This is also visually captured in Fig.~\ref{fig:indirect_vs_direct}, which shows that the indirect water footprint can take more than 40\% of the overall operational water footprint. For example, for \polaris{}, the indirect operational water footprint accounts for almost 47\% of the operational water footprint. 

\vspace{-1mm}

\takeaway{
Indirect operational water footprint can often be comparable to direct operational water footprint -- that is, \textit{the water consumed to generate electricity (besides water needed to cool the HPC system) must be taken into account toward overall water-optimized HPC system operations.} Unfortunately, the factors that affect these components show strong temporal and spatial variation. Favorable climate conditions for cooling the datacenter do not necessarily mean overall lower water consumption -- since water consumed during energy generation may become dominant.  
}

\vspace{-1.5mm}
\takeaway{
An implication for \textit{HPC facilities and city operators is that they should dynamically determine what fraction of total water goes where (``water capping'') when water is a constrained resource -- toward the cooling of the datacenter, or toward energy generation.} For example, when the weather conditions are less favorable (high WUE; more water for cooling is needed), the power grid should focus on generating energy from less water-intensive sources, possibly at the expense of carbon footprint. 
}

\begin{figure}[t]
    \centering
\includegraphics[scale=0.44]{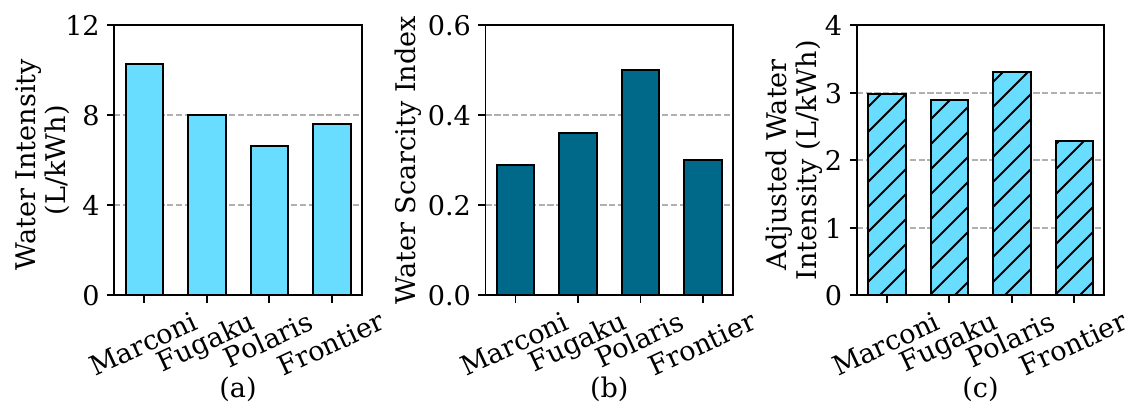}
\vspace{-8mm}
    \caption{(a) Annual average water intensities (water footprint) across different regions, (b) water scarcity index of different regions using AWARE-global data, and (c) the adjusted water intensity after combining the water intensity with the water scarcity index.}
    \label{fig:WSI}
\end{figure}

\begin{figure}[t]
    \centering
\includegraphics[scale=0.35]{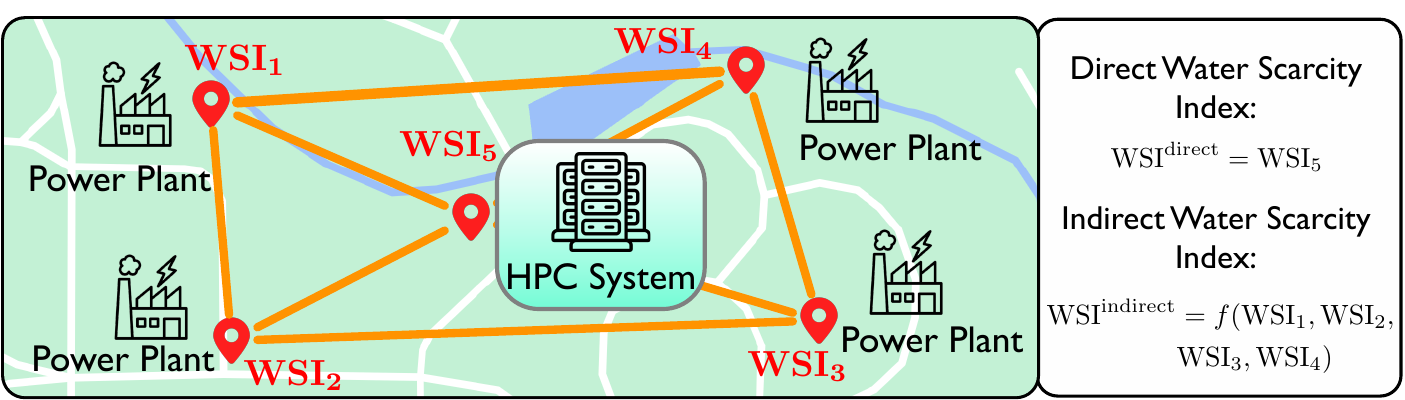}
\vspace{-3mm}
   \caption{Direct and indirect water scarcity index and its impact on overall water intensity calculation.}
   \vspace{-4mm}
    \label{fig:direct_indirect_wsi}
\end{figure}

\begin{figure*}[t]
    \centering
\includegraphics[scale=0.43]{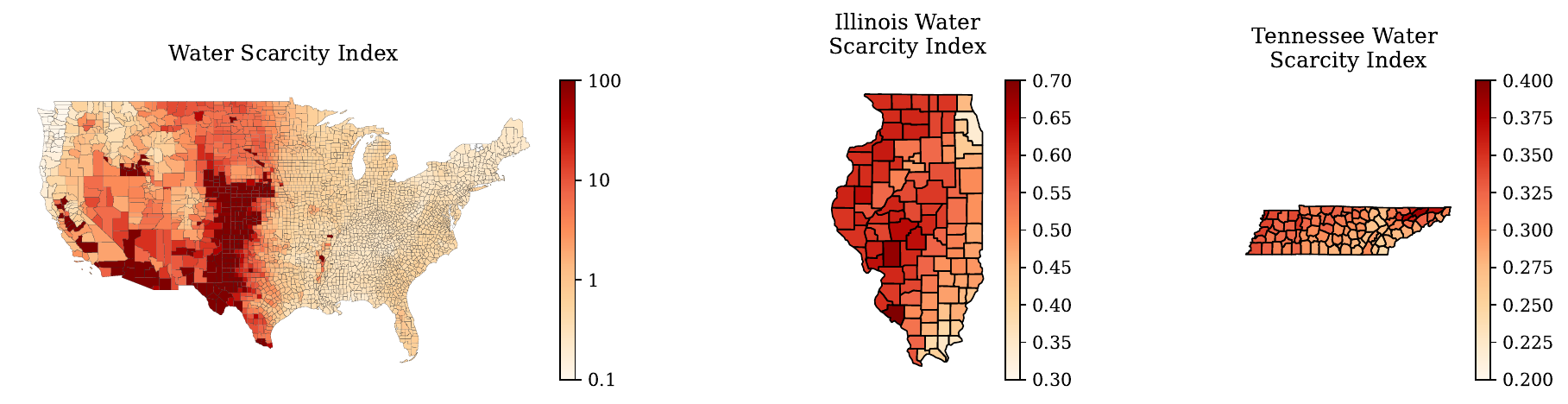}
    \caption{Direct and indirect WSIs exhibit significant variation for both Illinois and Tennessee, and throughout the USA.}
    \label{fig:diff_WSI}
\end{figure*}

\vspace{2mm}
\noindent\textbf{\large{Impact of Regional Water Scarcity.}} Next, we investigate the impact of the regional water scarcity index (WSI) on the overall water footprint. Intuitively, a larger water scarcity index implies that water is a scarce resource and hence, the adjusted or effective water consumption becomes higher. Fig.~\ref{fig:WSI} shows the normal water footprint/water intensity for four considered locations and then, the effective water footprint/water intensity after accounting for WSI. Notably, the relative ranking of the regions changes once WSI is considered. Even though, \polaris{} consumes the least water per kWh as such, because of the high water scarcity index of Chicago and nearby areas, its effective water intensity becomes the highest. 

Next, we demonstrate that the methodology for accounting for the WSI is rather challenging and requires careful consideration. This is because an HPC center can have multiple WSIs that affect the overall water footprint estimation -- the regional WSI of the datacenter’s location (direct WSI), and the regional WSI of the electricity generation site (indirect WSI). Interestingly, an HPC center may be receiving energy from multiple locations and hence, may need to consider multiple WSIs -- this is visually depicted in Fig.~\ref{fig:direct_indirect_wsi}. To further support this, we show that the WSI indeed changes significantly at the county level -- Fig.~\ref{fig:diff_WSI} shows the state-level WSI data for Illinois and Tennessee, and it also shows the overall water scarcity data for the USA (fine-grained WSI data is unavailable for non-US locations).

\vspace{1mm}

\takeaway{
As expected, the geographical dependence of the water scarcity index affects the effective water footprint of different HPC centers. Interestingly, the water scarcity index can vary significantly even at a kilometer scale, and hence, the operational water footprint is greatly impacted by which nearby power grids are being used for electricity generation. \textit{Hence, the HPC center operations should consider accounting for the water scarcity index of all nearby power grids, besides electricity cost and renewability of the energy mix (carbon intensity).}
}

\section{Energy, Water, and Carbon Footprint}
\label{sec:water_and_carbon}

In this section, we investigate the interplay between three important sustainability metrics: energy, water, and carbon.

\begin{figure}[t]
    \centering
\includegraphics[scale=0.4]{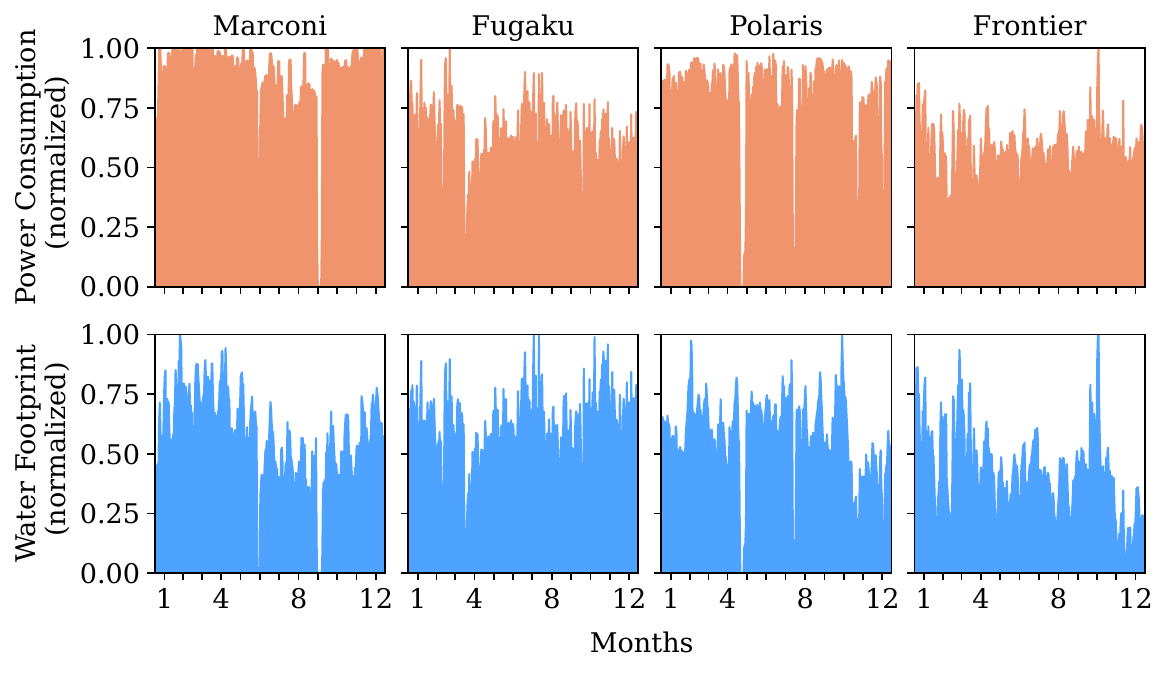}
    \caption{Temporal energy consumption (top) and water footprint (bottom) variations over one year for four HPC systems.}
    \label{fig:energy_and_water}
\end{figure}

It is reasonable to expect that the energy consumption of an HPC system is correlated with its water footprint. While this expectation is intuitive, it is not always necessarily true. Figure~\ref{fig:energy_and_water} shows the estimated energy consumption and the water footprint (operational) of four HPC systems over one year. Both metrics are normalized using min-max scaling, allowing us to compare relative variations within each system. There is a correlation in patterns, as expected, but they do not exactly align. This is because water footprint also depends on outside weather conditions (WUE), water consumption during energy generation (EWF), and the mix of energy sources (depending on the regional energy source availability). Unfortunately, these factors can have different temporal patterns even when the load on the system (energy consumption) is constant -- explaining the differences in patterns.

\takeaway{
\textit{As an implication, energy-aware HPC system operation does not necessarily mean water-optimal operation.} Existing popular research strategies that attempt to minimize energy consumption (e.g., workload shifting among HPC centers purely based on energy consumption) may still lead to disproportionately high water use if regional water constraints and patterns are not carefully considered.  
} 
\vspace{2mm}

\begin{figure}[t]
    \centering
\includegraphics[scale=0.46]{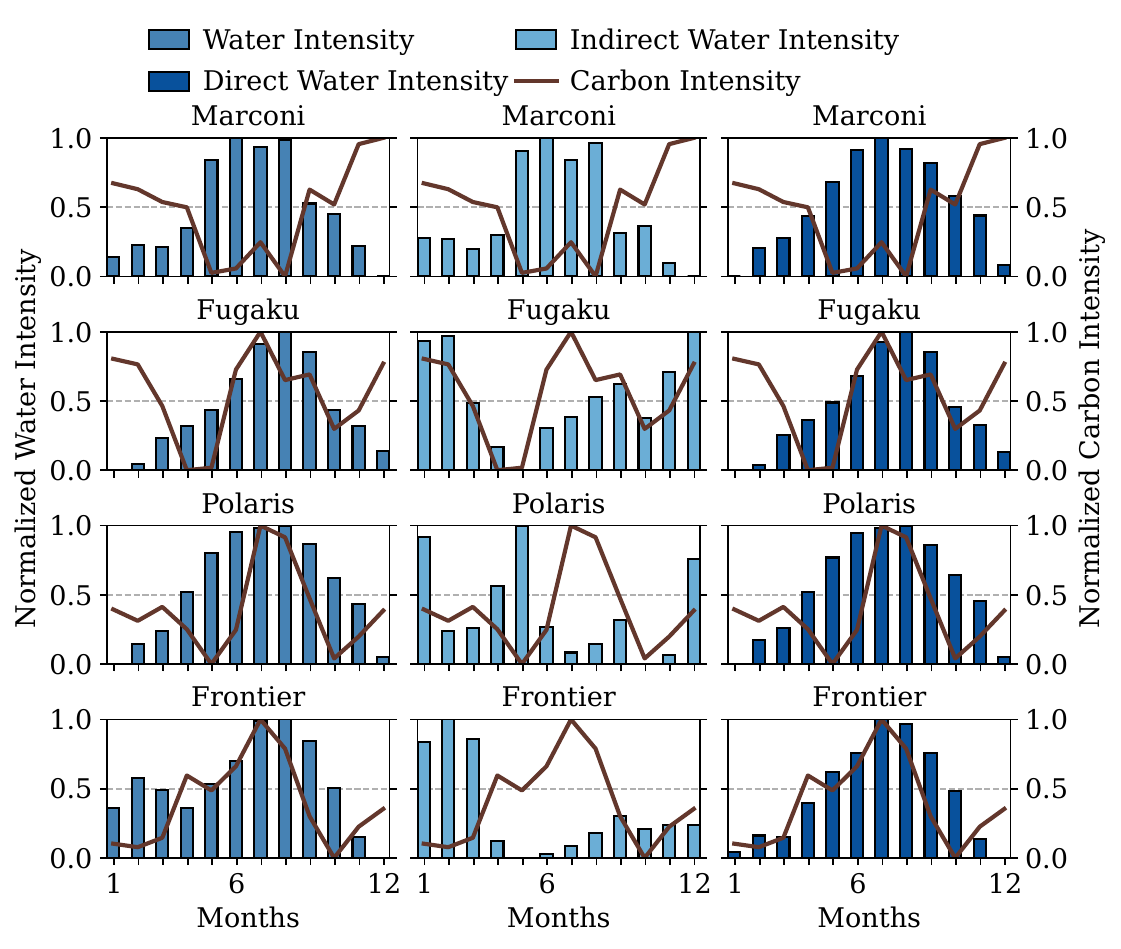}
    \caption{Carbon intensity can compete with overall water intensity, primarily through the indirect water footprint component. }
    \label{fig:ci_wi}
\end{figure}

Intuitively, one may expect a competing trade-off between water and carbon. In Fig.~\ref{fig:ci_wi}, we show the monthly variations of overall water intensity, carbon intensity, direct water intensity, and indirect water intensity for four evaluated HPC systems. We make two interesting observations. First, as expected, the overall water intensity is high during summer months due to the need for more water to cool the datacenter (higher direct water intensity) -- unfavorable outside weather conditions.  

Second, carbon and water trends have interesting interactions -- sometimes similar trends and sometimes competing trends. This is because the carbon intensity entirely depends on the renewable nature of the energy source (highly renewable energy sources have lower carbon intensity) and the temporal variation in the energy source mix. However, water intensity, particularly the indirect water intensity, depends on the EWF factor (water required to generate electricity). Even highly renewable energy sources (e.g., recall Fig~\ref{fig:diff_energy}) can have high water intensity -- leading to a high indirect water footprint despite a lower carbon footprint. This is the reason for competing water and carbon footprint trends in \marconi{} during the summer months.

\takeaway{
Fortunately, both major sustainability metrics (carbon and water footprint) do not always compete, but the future HPC system design should carefully consider both metrics explicitly -- \textit{depending upon the geographical location and temporal energy source mix, a carbon-friendly HPC system may result in a large indirect water footprint}. The decision-making process~\cite{hpcwire2025doe} for HPC and AI datacenters should explicitly have both synergistic and competitive interactions between carbon and water. 
}

\vspace{2mm}

In fact, to further demonstrate the above insight at a finer time-scale (hourly instead of monthly as shown in Fig.~\ref{fig:ci_wi}), we demonstrate that the ``optimal'' time to execute an application, even with a single HPC system, is different from carbon and water perspective.

\begin{figure}[t]
    \centering
\includegraphics[scale=0.41]{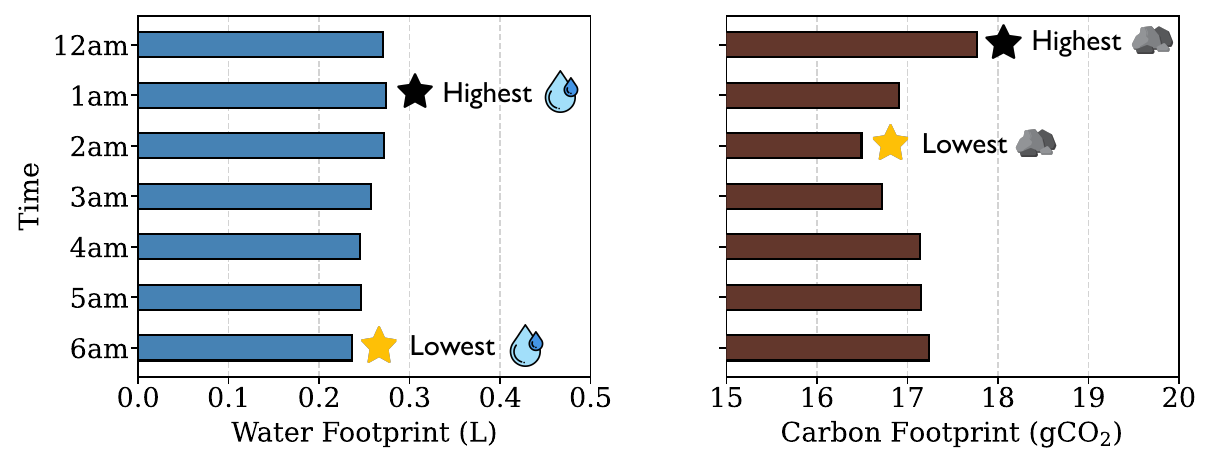}
    \caption{Ranking of potential application start times of application execution based on water and carbon impacts.}
    \label{fig:app}
\end{figure}

Fig.~\ref{fig:app} shows the results for an experiment we conducted on an Intel Xeon Platinum 8175 CPU and 384 GB of memory, to demonstrate the insight, assuming the rest of the HPC center characteristics are the same as Frontier. We executed miniAMR~\cite{vaughan2017evaluating}, a mini-application that performs stencil computations on a unit cube using adaptive mesh refinement. We evaluate the temporal impact, select seven potential start times, and compare their suitability in terms of environmental impact. We note that in all cases, as expected, the miniAMR consumes the same amount of energy. As shown in Fig.~\ref{fig:app} the most suitable times for carbon and water are different. This is because water and carbon intensities vary on an hourly timescale, exhibiting periodic troughs and peaks throughout the day. Consequently, the timing of HPC application execution plays a critical role in minimizing carbon and water footprints. 

\takeaway{
As such, programmers do not necessarily need to invent additional tools for optimizing water consumption -- because minimizing energy consumption achieves a similar effect from the programmers' side without measuring water consumption --, but new schedulers need to be developed if the HPC centers want to co-optimize for multiple sustainability metrics such as water, carbon, energy, etc. 
}
\vspace{2mm}

Next, we investigate the interaction between carbon and water footprint more in-depth by focusing on nuclear energy, which is touted to be carbon-friendly.
\section{Nuclear Reactor Powered HPC}
\label{sec:nuclear}

\begin{figure}[t]
    \centering
\includegraphics[scale=0.44]{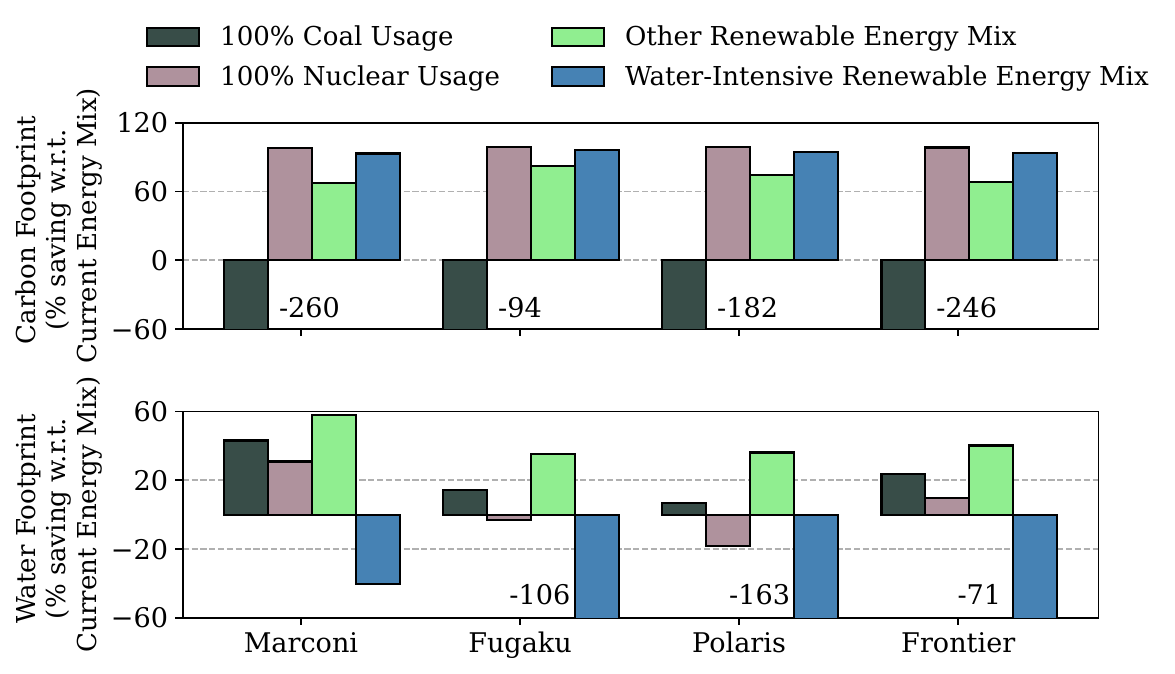}
    \caption{Impact of nuclear and other renewable energy sources on carbon and water footprint for different HPC systems.}
    \label{fig:nuclear}
\end{figure}

Our exploration is motivated by a recent emergent interest in nuclear energy and its carbon-friendliness. Amazon and Google are projected to develop small nuclear reactors to power the datacenters~\cite{GoogleKairosAgreement2024, amazon2024smr}. Unfortunately, the water footprint of nuclear energy is suspected to be high, but not well understood in the context of HPC systems. Therefore, in Fig.~\ref{fig:nuclear}, we quantify different scenarios in terms of both carbon footprint and water footprint: current energy source mix (normalization point), 100\% coal energy (non-carbon-friendly), 100\% nuclear energy, non-water-intensive 100\% renewable energy source (e.g. solar, wind), water-intensive 100\% renewable energy source (e.g. hydro).

Before drawing conclusions, we want to acknowledge the limitations and assumptions - we recognize that not all locations may have availability of these five considered energy sources, and building nuclear power plants to power HPC systems has additional considerations (cost, local regulations, health risks, etc.). 

We make two observations. First, as expected, in terms of carbon footprint, nuclear energy is on par with highly renewable energy sources -- and, yielding consistently over 80\% savings in carbon footprint over the current energy source mix at each HPC center. These are significant savings, especially given that highly non-renewable energy sources such as coal would result in an increase of more than 100\% in carbon footprint over the existing setup at all HPC centers. The second observation is more interesting. We find that, unlike carbon footprint, the water footprint of nuclear energy compared to the existing energy source mix is very location-dependent -- for \marconi{} and \frontier{}, it results in water savings, but for other systems, it may increase the water footprint. This is because of the water requirement during the energy generation process using nuclear reactors. 

Nuclear reactors use large volumes of water to condense steam and dissipate waste heat~\cite{macknick2011review}. A nuclear plant with a conventional wet cooling tower may consume 2.2-3.2~L/kWh. Even in the favorable setup where once-through cooling is used (drawing water from a river and returning most of it), the energy water factor is 0.5-1.5~L/kWh. Therefore, in water-stressed regions, nuclear power reactors may not be a suitable choice -- and, other forms of non-water-intensive energy sources (e.g., biomass) should be explored to power the HPC systems. Nevertheless, nuclear power is not necessarily the worst in terms of its water requirement. For example, water-intensive 100\% renewable energy sources such as hydro may result in more than 60\% increase in water footprint compared to the existing energy source setup.

\vspace{1mm}
\takeaway{
Small nuclear reactors are a promising option for mitigating the exploding energy needs of datacenters and their environmental carbon footprint. \textit{But, the water footprint of nuclear power reactors can be significant -- and, importantly, the impact is location-dependent. Naively employing nuclear reactors to power HPC centers, to mitigate energy and carbon footprint concerns, can be significantly sub-optimal depending upon the location.}
}
\vspace{3mm}
Finally, we acknowledge that even though nuclear reactors may appear to be a promising choice, nuclear power comes with several important concerns -- including regulation, health concerns, long-term impact, time required to build solutions, cost, etc. Many of these concerns are not as well explored yet, but given the increasing interest, nuclear-powered HPC should be studied more carefully. 

The water consumption itself, whether nuclear or not, may require more carefully thought-out regulations. Unlike carbon, water can have a ripple effect and it can cause effects in neighboring locations. For example, one country's extensive dam-building projects can provide hydroelectric power, but they significantly reduce downstream flow to other neighboring countries -- or, even impact their climate. In general, for river basins, upstream water withdrawal can jeopardize downstream water availability, potentially causing ecological problems. With an increasing water consumption of HPC centers, we hope that this study initiates a much-needed discussion and effort to make our HPC systems more water-aware.

\section{Discussion and Outlook}

In this section, we discuss aspects of \sol{} that are important and will benefit from more research in the future. 

\vspace{2mm}
\noindent\textit{Water in Datacenter Construction. } The data for construction-related water use (e.g., for concrete, steel, and other materials) is currently limited and site-specific. The common way is to maintain a database of Life Cycle Assessment for the datacenter construction. While it is an important aspect, the lack of open-source and standardized data makes it difficult to estimate or include in the modeling.

\vspace{2mm}
\noindent\textit{Water Withdrawal.} In addition to quantifying water consumption, \sol{} can also estimate water withdrawal with the parameters in Table~\ref{table:water_with}. Water withdrawal is derived from discharge water, water consumption, and water reuse. (1) \emph{Discharge Water.} It represents the portion of withdrawn water returned to the environment~\cite{zhang2024data_center_water_usage}. The impact of discharge depends on both the location of the outfall ($L_k$) and the hazard level of the discharged pollutants ($P_j$) in the water. For example, wetlands provide natural purification benefits, and rivers are treated as neutral receivers. In addition, pollutant hazards, such as biological oxygen demand (BOD), chemical oxygen demand (COD), and heavy metals, are applied to scale the discharge accordingly~\cite{thomsen2016wastewater,mytton2021data}. These parameters allow us to normalize water discharge into a comparable metric that accounts for both environmental context and pollutant severity. The adjusted water discharge can be estimated with the actual reported water discharge $W_{\text{discharge}}^{\text{actual}}$. (2) \emph{Water Reuse.} Water reuse captures the fraction of discharged water that is recycled within the system~\cite{spindler2024circular_water_sustainable_data_centres}. Formally, we define the water reuse as the product of the discharge water and the water reuse rate $\rho$. (3). \emph{Potable and Non-potable Water.} Withdrawn water can be classified into potable and non-potable water, reflecting the sources of water resources~\cite{mytton2021data}. $\beta_{\text{potable}}$ and $\beta_{\text{non-potable}}$ in Table~\ref{table:water_with} define the percentage of potable and non-potable water in total water withdrawal, respectively. Similar to water discharge, different water sources can have varying resource scarcity factors ($S_{\text{potable}}$, $S_{\text{non-potable}}$), which range from 0 to 1, with higher values indicating more limited resources. Although water withdrawal is not presented in our results due to the lack of a standardized method for accounting, we provide a modeling methodology that can be used and enhanced by others. 

\begin{table}
\centering
\caption{Parameters for water withdrawal.}
\scalebox{0.7}{
\begin{tabular}{ccc}
\toprule
\textbf{Parameter} & \textbf{Description}& \textbf{Data Range} \\ 
\midrule
\midrule
$W_{\text{discharge}}^{\text{actual}}$ &  \makecell[c]{Reported discharge water footprint}
&  \makecell[c]{Vary across systems}\\
\midrule
$L_k$ &  \makecell[c]{Outfall location factor}
&  \makecell[c]{Vary across HPC locations} 
\\
\midrule
$P_j$&\makecell[c]{Pollutant hazard factor} &\makecell[c]{Vary across pollutants} \\
\midrule

$\rho$ &  \makecell[c]{Water reuse rate} 
&  \makecell[c]{0\%-100\%} \\
\midrule
$\beta_{\text{potable}}$/$\beta_{\text{non-potable}}$&\makecell[c]{Percentage of potable/non-potable water} &\makecell[c]{0\%-100\%} \\
\midrule
$S_{\text{potable}}$/$S_{\text{non-potable}}$&\makecell[c]{Scarcity factor (potable / non-potable)} &\makecell[c]{Vary across water sources} \\
\bottomrule
\end{tabular}}
\label{table:water_with}
\end{table}

\vspace{2mm}
\noindent\textit{Embodied Water Consideration.} \sol{} accounts for embodied water during hardware manufacturing to make the modeling more comprehensive and accurate. Modeling this component allows chip designers to estimate and reduce the embodied water consumption during hardware manufacturing, and system procurement teams to consider water consumption during hardware manufacturing. Additionally, this component is critical for accurate comparison across different HPC systems with various hardware types and upgrade cycles, and hence, should continue to be included in future studies.

\vspace{2mm}
\noindent\textit{Broader and Future Usages of \sol{}.} Most popular sustainability frameworks are carbon-focused; naturally, these tools, such as ACT~\cite{gupta2022act} and CarbonTracker~\cite{anthony2020carbontracker}, focus exclusively on carbon footprint and do not account for water usage. In contrast, \sol{} is the first framework to enable HPC system procurement and researchers to understand and quantify the impact of water consumption (operational decisions about energy mix, site selection, dynamically determining water allocation, and energy and water tension for job scheduling). Here are some examples that we hope the researchers can build upon in the future:

\vspace{2mm}
(a) \emph{Co-optimization of multiple sustainability metrics.} With the increasing societal impact of high water consumption in our HPC centers, water consumption will become a part of the multi‐objective optimizer. \sol{} can enable optimization techniques that assign adjustable weights to energy, carbon, and water metrics.

\vspace{2mm}
(b) \emph{Water footprint estimation for supercomputers.} \sol{} is not restricted to only the systems evaluated in the paper. It can be used for other HPC systems, including Aurora~\cite{Aurora2025} and El Capitan~\cite{ElCapitan2024}, with available or approximated parameters used in Table~\ref{table:tool}. While it is not a central focus of \sol{} to provide the ranking of HPC systems based on their water footprint, we hope it can inspire future research in this direction to make HPC system design and operations water-aware (e.g., Water500 ranking similar to performance-based rankings).

\section{Related Work}
\label{sec:related}

\vspace{3mm}
\noindent\textbf{Sustainability in HPC Systems. }The rapid growth of HPC has raised significant environmental concerns. Numerous studies have focused on analyzing and reducing the energy consumption and carbon footprint of HPC systems~\cite{benhari2024green,solorzano2024toward,li2023toward,Chadha2023,georgiou2015scheduler,saravanan2014performance,antici2023pm100,chu2024generic}. For example, a study~\cite{Chadha2023} emphasizes that achieving sustainability requires a holistic approach spanning hardware design, system architecture, lifecycle management, and user incentives to reduce the carbon footprint of HPC. Another work proposes a carbon footprint analysis framework~\cite{li2023toward} that characterizes carbon emissions in HPC systems and offers insights by addressing research questions. However, the water footprint remains largely overlooked. Our paper aims to fill that gap by addressing the current lack of understanding and analysis of the water footprint in the HPC community.

\vspace{3mm}
\noindent\textbf{Sustainability Modeling.} Recent computing sustainability efforts have developed frameworks~\cite{li2023making,gupta2022act,gupta2021chasing,Acun2023,basu2025forgetmenot,elgamal2025modeling,sudarshan2024eco} to quantify and optimize environmental impact (carbon, water, PFAS). Tools like the Architectural Carbon Modeling Tool (ACT)~\cite{gupta2022act} provide detailed estimates of carbon emissions from hardware, promoting carbon-aware design. The Carbon Explorer~\cite{Acun2023} for datacenters models how decisions in system configuration impact carbon emissions. ECO-CHIP~\cite{sudarshan2024eco} targets VLSI for estimating both embodied and operational carbon footprints of chiplet-based heterogeneous integration. A preliminary, water-related study, AI-thirsty~\cite{li2023making} provides a basic model for quantifying water footprints in datacenters, which motivates our work.

\vspace{3mm}
\noindent\textbf{Water-aware Computing. }Water usage in computing systems has become crucial, and system researchers have begun quantifying and reducing the water footprint~\cite{siddik2021environmental,li2023making,jiang2025waterwise,mohammad2018wace,alissa2025using,wu2025not}. A spatially detailed analysis~\cite{siddik2021environmental} of US datacenters demonstrates that strategic placement of datacenters could significantly reduce both water and carbon footprints, particularly by avoiding water-stressed regions. The WACE framework~\cite{mohammad2018wace} reduces water footprints in geographically distributed datacenters by dynamically scheduling jobs based on spatial and temporal variations, incurring only minor increases in job delays. The WaterWise framework~\cite{jiang2025waterwise} addresses trade-offs between carbon and water footprints in cloud computing by co-optimizing job scheduling across geographically distributed datacenters. However, all these works lack water analysis of the large-scale systems and are not related to HPC. By contrast, our work analyzes water footprints of real-world supercomputers from the TOP500 list, providing insightful takeaways to set up a comprehensive methodology for water quantification. 

\section{Conclusion}
\label{sec:conclusion}
In this paper, we have conducted a comprehensive analysis of the water footprint for four HPC systems. We develop \sol{}, the first tool to quantify both operational and embodied water footprints of HPC systems and examine how water intensity influences overall water consumption in these systems. Additionally, we investigate the impact of water scarcity and compare the water footprint with the carbon footprint, highlighting the inherent trade-offs between achieving low-carbon operations and ensuring sustainable water usage. Finally, we assess the implications of cutting-edge nuclear-powered HPC scenarios, bringing out the advantages and warnings for nuclear-powered HPC. We hope \sol{} and the data can be the foundation for future research aimed at developing water-aware computing strategies for HPC systems.

\vspace{3mm}
\noindent\textbf{Acknowledgment}. We thank the reviewers for their constructive and insightful feedback. This work
is supported by NSF Awards 1910601, 2124897, and Northeastern University.

\bibliographystyle{ACM-Reference-Format}
\bibliography{refs}

\end{document}